\documentclass[twocolumn]{article}
\usepackage{a4wide,graphicx}%amsmath,amsfonts,amssymb}
\usepackage{natbib}

\begin{document}

\title{A MODEL OF SEQUENCE DEPENDENT PROTEIN DIFFUSION ALONG DNA.}

\author{Maria Barbi $\,\,{^\dagger}{^{\P}}$, Christophe Place$^\#$, Vladislav
Popkov${^{\S}} {^{||}}$, Mario Salerno${^\dagger}$}

%\date{\today}

{
%\twocolumn[

\maketitle

${^\dagger}$~Dipartimento di Fisica ``E.R. Caianiello'' and INFM,
Universit{\`a} di Salerno, Baronissi (SA), Italy; $^\#$~Laboratoire de
Physique, CNRS-UMR 5672, {\'E}cole Normale Sup{\'e}rieure de Lyon,
Lyon, France; $^{\S}$~Institut f\"{u}r Festk\"{o}rperforschung,
Forschung\-szentrum J\"{u}lich GmbH, J\"{u}lich, Germany;
$^{||}$~Institute for Low Temperature Physics, Kharkov, Ukraine; and
$^{\P}$~Laboratoire de Physique Theorique des Liquides, Universit{\'e}
Pierre et Marie Curie, case courrier 121, 4 Place Jussieu - 75252
Paris cedex 05, France; ~e-mail:~{barbi@lptl.jussieu.fr} .

\vspace{3mm}

KEYWORDS: sliding, promoter search, anomalous diffusion, DNA-RNAP
interaction, dynamical models

\vspace{3mm}

{\begin{abstract} We introduce a probabilistic model for protein
sliding motion along DNA during the search of a target sequence. The
model accounts for possible effects due to sequence-dependent
interaction between the nonspecific DNA and the protein.  As an
example, we focus on T7 RNA-polymerase and exploit the available
information about its interaction at the promoter site in order to
investigate the influence of bacteriophage T7 DNA sequence on the
dynamics of the sliding process. Hydrogen bonds in the major groove
are used as the main sequence-dependent interaction between
RNA-polymerase and DNA.  The resulting dynamical properties and the
possibility of an experimental verification are discussed in details.
We show that, while at large times the process reaches a pure
diffusive regime, it initially displays a sub-diffusive behavior.  The
subdiffusive regime can lasts sufficiently long to be of biological
interest.
\end{abstract}
}

\vspace{1cm}

%]  %end one-colomn-part of the command \twocolumn

%{\footnotesize \tableofcontents}

%%%%%%%%%%%%%%%%%%%%%%%%%%%%%%%%%%%%%%%%%%%%%%%%%%%%%%%%%%%%%%%%%%%%

\section*{Introduction}
\label{intro}

The way by which proteins can find their target sites along a DNA
chain represents a puzzling problem. In many cases, the reaction
rate has been demonstrated to be faster than diffusion controlled
\citep{Rig70,Ber81,Rei91,Sur96}. Nonspecific sliding along the
DNA has been proposed to be the main mechanism for faster search of
the specific site on DNA
\citep{Par82a,Par82b,Sin87,Ric88,Kab93,Gut94,Scu98,Gut99,Shi99,Har99}.
Nevertheless, a precise experimental determination of the statistical
law characterizing the diffusion motion of protein along DNA during
the specific site search is presently lacking. It is believed that
during the sliding motion, the activation barrier for the
translocation of the protein to continuous nonspecific positions is
high enough to randomize the protein motion through collisions with
the solvent water, but appropriately small compared to the thermal
energy, in order to allow the protein to move \citep{von89}. This has
induced some authors to propose a model where protein freely slides
along DNA under the effect of the thermal fluctuations without any
sequence dependent interaction, i.e., the DNA is seen as an
homogeneous cylinder on which the protein can diffuse until the
specific site is reached \citep{von89,von96,Par82b}. During sliding,
however, the protein must be able to distinguish the specific region
from nonspecific DNA so that a recognition mechanism must be
involved.  To this regard, the possibility that sliding could imply
sequence dependent protein-DNA interaction is rather reasonable.

The aim of the present paper is to investigate this idea in the
context of a simple probabilistic model for RNA-polymerase (RNAP)
sliding along DNA, which accounts for the sequence-dependent
interaction between the nonspecific DNA and the enzyme. As an
illustrative example we consider the case of the T7 RNA-polymerase
sliding on the bacteriophage T7 DNA \citep{Dun83}. Although the
results of the paper are likely to be valid also for other enzymes,
the T7 RNAP has several advantages which are suitable for our
modelling. In particular, we mention the simplicity of the enzyme, (it
is a small enzyme ($100\, kDa$) composed of only of one unit, and
recognizes a single asymmetric region on the DNA), and the
availability of high resolution crystal structure data both for the
RNAP alone and for the RNAP bound to his promoter
\citep{Jer98,Che99,Che99b}. In contrast to more complicated enzymes
such as lac repressor \citep{Win81}, restriction endonuclease (EcoRI
\citep{Jac82,Ehb85}, EcoRV \citep{Dow90,Sta94}), methyl transferase
(EcoRI \citep{Sur96}), {\it E. coli} RNA polymerase \citep{Par82b}, etc., no
direct evidence of diffusive sliding motion has been presented for T7
RNAP. However, the fact that the enzyme is able to locate his
promoters inside about $40000$ base pairs DNA during a time much
shorter than what a three dimensional search would require
\citep{End00}, strongly suggests a sliding mechanism also in this
case. In our model we assume therefore that T7 RNAP proceeds by
sliding during the promoter search. The model is based on the idea
that the RNAP needs to ``read'' the underlying sequence during sliding
in order to test whether special ``signals'' associated with the
promoter are present, i.e., a sequence-dependent interaction should be
at work during the search. This means that the DNA sequence can
influence the dynamics of the polymerase also far from the
promoter. In this sense, the stop at the promoter should be the
extreme effect of a complex dynamics, i.e., the RNAP should follow a
noise-influenced, sequence-dependent motion that includes the
possibility of slowing down, pauses and stops. From this point of view
the usual assumption of a standard random walk of the RNAP along DNA
\citep{Ber81,Kab93,Har99,Gut99,Sta00} appears inadequate.

To investigate the possibility of a sequence-dependent diffusion
motion of the RNAP along the DNA, we define a base sequence energy
landscape from which hopping rates of the enzyme on the DNA (view as a
discrete inhomogeneous lattice) can be deduced. Since only limited
experimental knowledge exists about nonspecific DNA-protein
interaction, we shall use information about sequence dependent
RNAP-DNA interaction inside the promoter region and extrapolate it to
nonspecific regions. The diffusive motion of the RNAP is then studied
by Monte-Carlo simulations of the probabilistic process on the
landscape energy both in absence and in presence of thresholds which
define different rules for the hopping motion. As a result we show
that while at large times the process reaches a pure diffusive regime,
at the initial stage it displays a sub-diffusive behavior. It is
remarkable that the anomalous diffusion regime can last for time large
enough to be observable in single molecule experiments similar to
those that have permitted to visualize sliding for the {\it E. coli} RNAP
\citep{Kab93,Har99,Gut99}. Singule molecule experiments on T7 RNAP are
indeed underway in several laboratories
\citep{Heslot,Baumann,Place}. We remark that base sequence induced
dynamics along DNA was also considered in Ref. \citep{sa91,sa95} in
connection with a nonlinear model of DNA, and in Ref.~\citep{Jul97b}
in connection with the RNAP motion during the transcription process.

The paper is organized as follows: in Section \ref{methods} we use
some known data on the T7 RNAP-promoter complex to introduce a
sequence dependent model for the RNAP-DNA nonspecific interaction. An
energy landscape with minima corresponding to the recognition sequence
is constructed. We then introduce four possible models for the RNAP
diffusive motion along the DNA by using the sequence induced energy
landscape and its modification as the inclusion of energy thresholds,
which allow to describe different possible reading mechanisms. The
rate of translocation to the neighboring sites is constructed from the
energy landscapes (for the different models) by means of the Arrhenius
law. In Section \ref{results} we use Monte-Carlo simulations to study
in detail the different dynamical regimes of our models. Finally, we
discuss in section \ref{comments} the limits of our analysis and the
possibility to check the results with experiments, so as to verify if
the inferred mechanism actually corresponds to the real one. Then we
draw our Conclusions.

%%%%%%%%%%%%%%%%%%%%%%%%%%%%%%%%%%%%%%%%%%%%%%%%%%%%%%%%%%%%%%%%%%%%%
\section{Methods: experimental data and theoretical model}
\label{methods}
\subsection{T7 RNAP - DNA interaction and promoter recognition}

The stability of the RNAP-DNA nonspecific complex is mainly due
to electrostatic interaction with the backbone phosphate of DNA
\citep{von96}  and to the entropic release of cations
\citep{deH77,Sid01,Sin88}. For specific interaction, while ionic
effect could still be present \citep{Rec77}, the major stabilization
effect arises from the release of water molecules \citep{Sid01}. The
presence of a layer of water between protein and DNA in nonspecific
complex weakens the specific interaction. This suggests that a
continuous variation between specific and nonspecific binding exists
\citep{Jel94}; the transition from nonspecific to specific complex
can be induced by conformational changes of the proteins \citep{Spo94}.

Besides these stabilizing factors, sequence-dependent interaction
allows the RNA polymerase to test the DNA during the promoter search
\citep{travers}. Experimental data on endonuclease EcoRI show that
pausing of the protein during sliding occurs at sites which resemble
the specific sequence \citep{Jel94}. Thus, the nonspecific
``reading'' should be of the same nature as the specific
recognition\footnote{Also remark that, for the case of CRP protein,
nonspecific binding have been proposed to mimics the specific,
c-Amp-dependent binding \citep{Kat93}, this confirming the hypothesis
of a continuity between nonspecific and specific recognition
interaction.}. From this observation one can deduce that at least some
of the different kinds of interaction observed in the specific complex
could be already present during sliding, and might be used in the
recognition mechanism. This hypothesis can be interesting also if the
actual reading mechanism is not exactly the same but a similar type;
the study of the consequent dynamics may help, from a general point of
view, in understanding which kind of sequence-dependent interaction is
compatible with the experimental data.

The first point to address, in order to have a suitable description of
the promoter search dynamics, is therefore to determine which
sequence-dependent interaction is responsible of the promoter
recognition by the T7 RNAP. Experimental results seem to indicate, as
we will now discuss, that a specific set of hydrogen bonds on the 5
bps sequence GAGTC represents the main recognition core. We will
therefore use this set of bonds as the main recognition tool in our
model.  Biochemical and structural analysis gives a very precise
information on the principle of promoter recognition
\citep{McA97,Sou97} and on the polymerase-promoter specific complex
for the case of bacteriophage T7 \citep{Che99}. T7 RNAP recognizes a
23 bps promoter, that extends from -17 to +6 relatively to the
initiation site and consists of two functional domains
\citep{McA97}. It is reasonable to assume that the initiation domain,
extending from -4 to +6, does not interfere directly in the promoter
search: a measure of the dissociation constant with small
oligonucleotides carrying truncated promoter have shown indeed that
these base pairs do not participate in the promoter recognition
\citep{Ujv97}. Let us then consider the binding domain (from -17 to -5).

\begin{figure}[htbp]
  \begin{center}
    \includegraphics[width=0.48\textwidth]{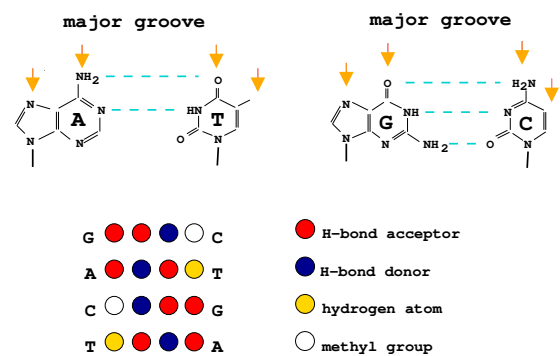}
    \caption{The positions of all the possible major groove
    interacting sites where base-pairs can make hydrogen bonds {\em
    (top)} and the corresponding base-pair patterns {\em
    (bottom)}. Blue and red disks indicates the hydrogen donor and
    acceptor DNA groups respectively. White positions correspond to
    hydrogen atoms and yellow ones to methyl groups. Each base-pair is
    associated with a different $1 \times 4$ pattern.}
    \label{fig:groove1}
  \end{center}
\end{figure}

Different biochemical studies, together with a recent crystallographic
analysis, contribute to the determination of the most relevant base
pairs in this region. On one hand, a hierarchy of base pairs
preferences was determined by single points mutations in the promoter
\citep{Cha87,Dia93,Imb00}. These studies have shown lower
sequence-sensitivity of the region -17 to -12: the specificity arises
from bases -11 to -5, being more stringent on bases -7 to -9.  The
identification of the functional group of the DNA involved in those
potential contacts shows that direct contact in the recognition region
-11 to -5 arise mostly through the major groove of a double strand
promoter \citep{Sch95,Li96}. On the other hand, the crystal structure
of T7 RNAP bound to its promoter \citep{Che99} is consistent with
these biochemical studies \citep{Imb00} and draws a structural picture
of the T7 RNAP promoter interaction\footnote{See Figs.~1 and 2 in
Ref.~\citep{Che99} for a clear representation of the whole set of
interactions.}.  In particular, a set of sequence-specific bonds
between protein side chains and bases in the major groove arise in the
region -11 to -7, via the formation of hydrogen bonds with the
appropriate acceptor or donor chemical groups in the base pairs sides
(See Fig.~\ref{fig:groove1} and Fig.~\ref{fig:groove2}).

 \begin{figure}[htbp]
  \begin{center}
    \includegraphics[width=0.48\textwidth]{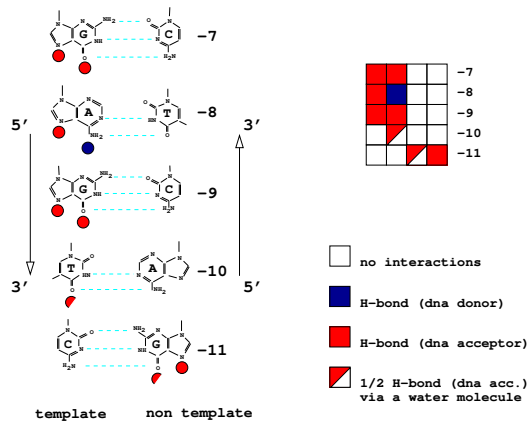}
    \caption{A sketch of the DNA interaction sites at the
    promoter, where hydrogen bonds with corresponding RNAP chemical
    groups are made. Blue and red disks indicate the hydrogen donor
    and acceptor DNA groups respectively; the two half disks
    correspond to a couple of sites that could share a water mediated
    hydrogen bond. On the right, the corresponding $5 \times 4$
    pattern that RNAP recognizes.}
    \label{fig:groove2}
  \end{center}
\end{figure}

We remark that the previously mentioned kinetic studies suggest that
base pairs -5 and -6 also can contribute to the recognition mechanism
\citep{Li96}: these contacts are probably lost once the open complex
is formed, so that the mentioned crystallographic analysis does not
show them.  Anyway, the base specificity appears to be less stringent
for these two contacts too \citep{Li96}. Because of their weak
specificity, we will neglect these two interacting base pairs, and
focus here just on the hydrogen bond mediated interaction arising on
bases -11 to -7, that has the strongest sensitivity to the base
pairs. The question addressed will be therefore how the specific
interaction of this 5 bps region influences the polymerase motion.

Hydrogen bond acceptors and donors are regularly positioned on the
promoter major groove: the DNA geometry is in fact such that each of
the four different base pairs exposes four possible major groove
interacting sites as depicted in Fig.~\ref{fig:groove1}
\citep{See76}. These sites can be either H-bond acceptors and
donors, or sites where a hydrogen atom or a methyl group are
present. In the latter case they do not bond directly to polymerase
(at least at the promoter site). Fig.~\ref{fig:groove2} depicts the
H-bonds actually made between polymerase and DNA at the promoter, as
revealed by the crystallographic analysis. The two semicircles in the
left part of Fig.~\ref{fig:groove2} and their correspondent positions
on the right pattern refer to the presence of a hydrogen bond which
is shared between two DNA sites through a water molecule
\citep{Che99}.

We shall assume that, in each position along DNA, the RNAP ``tries''
to make the same set of hydrogen bonds as at the promoter, testing in
this way the underlying sequence. It is convenient to represent the
RNAP by a recognition matrix able to match its target sequence, i.e.,
containing the pattern of active chemical groups that allows for the
best binding at the promoter. We suppose therefore that each position
along DNA will have a certain number of made ({\em matches}) and
unmade ({\em mismatches}) hydrogen bonds with the polymerase
recognition matrix.  For simplicity, we will represent the recognition
pattern directly in terms of its corresponding binding sites on
DNA\footnote{This is also a way to remind that we actually do not
include in the model all the possible polymerase-DNA nonspecific
bonds, but only those that are made at the promoter, for which an
experimental evidence is available.}.

One expects that each match will stabilize the complex, while
mismatches will act as to destabilize RNAP, that will tend
therefore to move away from the ``wrong'' positions
\citep{von89,von96}.  For each position $n$ along the chain we
define an energy $E(n)$, simply by counting the number of matches and
mismatches, and adding a corresponding negative or positive amount of
energy, respectively (empty sites in the recognition matrix do not
contribute to the energy).  Interacting sites corresponding
to the semicircles in Fig.~\ref{fig:groove2} are evaluated in a first
approximation as half hydrogen bonds everywhere along the chain.

Formally, the energy is defined by denoting by $+1,-1,0$ respectively
the acceptor, donor, and noninteracting DNA sites.  The DNA sequence
is then represented as a list of vectors, $
...b_{n-1}, b_{n}, b_{n+1}...$, where
\[
b_{n}=\left\{
\begin{array}{l}
(1,-1,1,0)^{T}  \,\,\,\mbox{for base A}\\
(0,1,-1,1)^{T}  \,\,\,\mbox{for base T}\\
(1,1,-1,0)^{T}  \,\,\,\mbox{for base G}\\
(0,-1,1,1)^{T}  \,\,\,\mbox{for base C}
\end{array}
\right.\] The polymerase acts, at position $n$, on the sequence
of 5 bases that is represented by the $4\times5$ matrix
$D_{n}=$($b_{n},b_{n+1,}b_{n+2},b_{n+3},b_{n+4})$. The consensus
sequence GAGTC at the promoter site corresponds therefore to the matrix
\[
D_{n}=\left(
\begin{array}{rrrrr}
1 & 1 & 1 & 0 & 0 \\
1 & -1 & 1 & 1 & -1 \\
-1 & 1 & -1 & -1 & 1 \\
0 & 0 & 0 & 1 & 1
\end{array}
\right).
\]

We then define a $4 \times 5$ recognition matrix $R(i,j)$, corresponding to
Fig.~\ref{fig:groove2},
\[ R = \left(
\begin{array}{rrrr}
1 & 1 & 0 & 0 \\
1 & -1 & 0 & 0 \\
1 & 1 & 0 & 0 \\
0 & 1/2 & 0 & 0 \\
0 & 0 & 1/2 & 1
\end{array}\right)
\]
where the factors $1/2$ have been introduced in order to reproduce the
shared hydrogen bond, previously mentioned.
With this notation, the interaction energy can be written simply as
\begin{equation}
  \label{eq:energy} E(n)= \, - \, {\mathcal \epsilon}\,\, tr ( R \cdot
  D_n )
\end{equation}
where the dot $\cdot$ denotes the usual matrix multiplication and $tr$
is the trace.  Minima correspond to the complete matching and thus to
the recognition sequence GAGTC. Each positive or negative contribution
to the energy, $\mathcal \epsilon$, is equal to a hydrogen bond
energy. Note that the mobility of RNAPs dramatically depends on
${\mathcal \epsilon} / k_B T$.  At room temperature, $k_BT$ is about
0.025 eV (or $RT$= 0.6 kcal/mol, $R=N_a k_B$); the energy barriers
must be smaller in order to allow the RNAP to move and reach the
promoter site\footnote{This may seem not consistent with the usual
measured strength of chemical hydrogen bonds, which normally
corresponds to a few kcal/mol {\citep{Stryer,Voet}}.  The distance and
orientation of the hydrogen bonds, however, together with their net
energetics due to the interaction with the solvent, may be responsible
of a relevant lowering of the interaction energy.}.  Since there are
no direct measurements of the interaction energies during sliding and
it is difficult to make an estimate of the involved hydrogen bond
strength, we shall use ${\mathcal \epsilon} / k_B T$ as a free
parameter.  The resulting energy $E(n)$ defines an irregular landscape
on which the RNAP can move as it will be discussed in the next
subsection.

%%%%%%%%%%%%%%%%%%%%%%%%%%%%%%%%%%%%%%%%%%%%%%%%%%%%%%%%%%%%%%%%%%%%%%%
\subsection{Sequence dependent RNAP diffusive model}

We shall introduce in this subsection four versions of the model
describing different mechanisms of the fundamental translocation step
in the enzyme motion.  The length of hydrogen bonds (up to $3.5$ {\AA}
in DNA-protein interaction {\citep{Nad99}}) can roughly reach the same
order of magnitude as the distance between base pairs ($3.4$
{\AA}). Therefore, the RNAP may eventually shift directly from one
position to the next one without activation energy for the one step
process. On the contrary, if the RNAP has to disrupt partially or
completely the hydrogen bonds on one site before moving to the next
position, it has to overcome an additional activation barrier.

Furthermore, RNAP could have some internal flexibility allowing for
conformational changes, eventually depending on the local degree of
stability: i.e., it is possible that, if too many mismatches are
found, RNAP should undergo a conformational change from a ``reading''
mode to a ``sliding'' mode, where no hydrogen bonds are effectively
made \citep{von96}. In this case, one has a sort of two-states model,
for which, if the total energy $E(n)$ is over a threshold $E_t$, the
system passes to a different state of constant energy $E_{sl}$ where
RNAP can freely slide.

\begin{figure}[htbp]
  \begin{center}
    \begin{tabular}{cc}
      \includegraphics[width=0.24\textwidth]{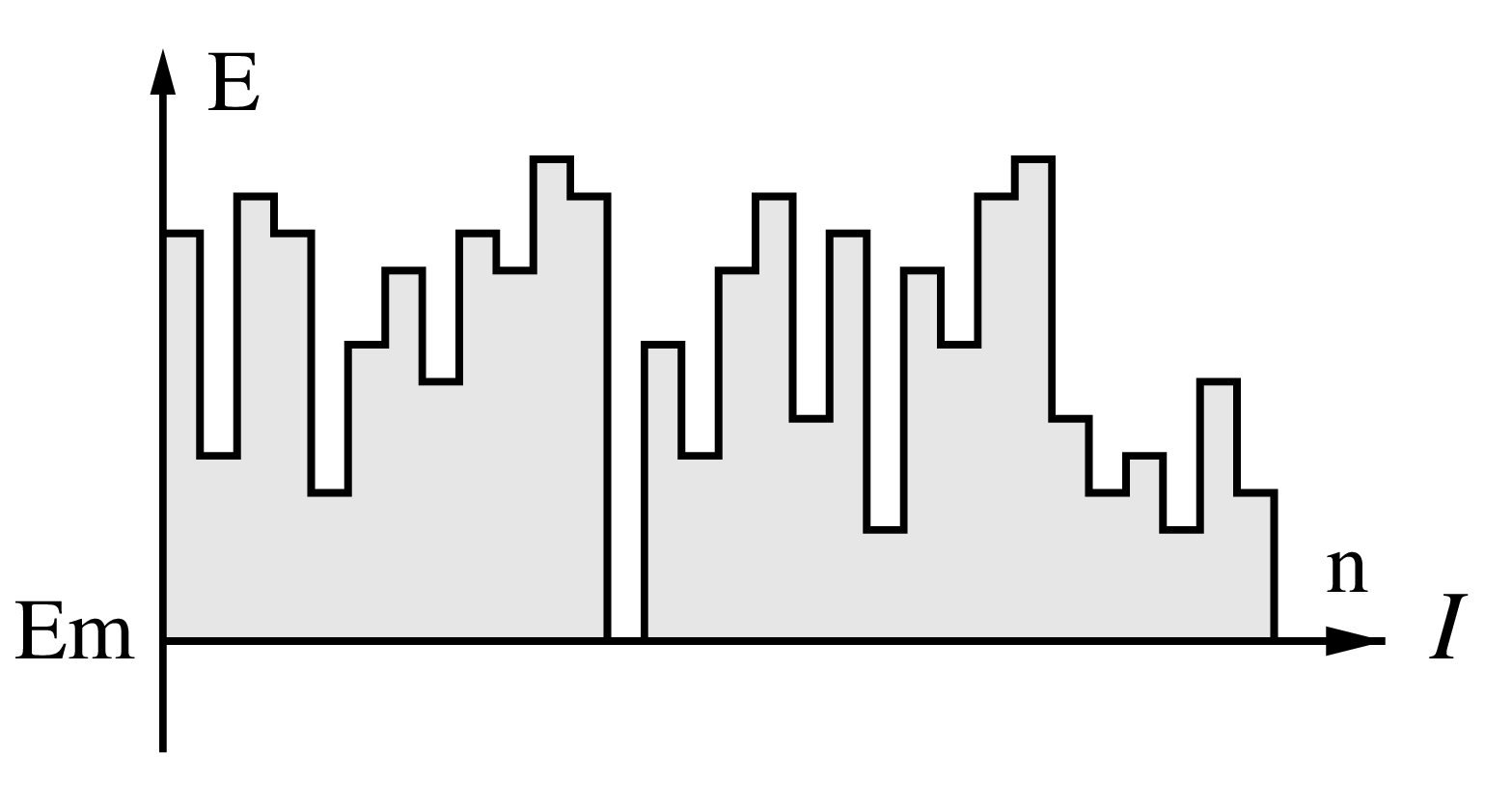} &
      \includegraphics[width=0.24\textwidth]{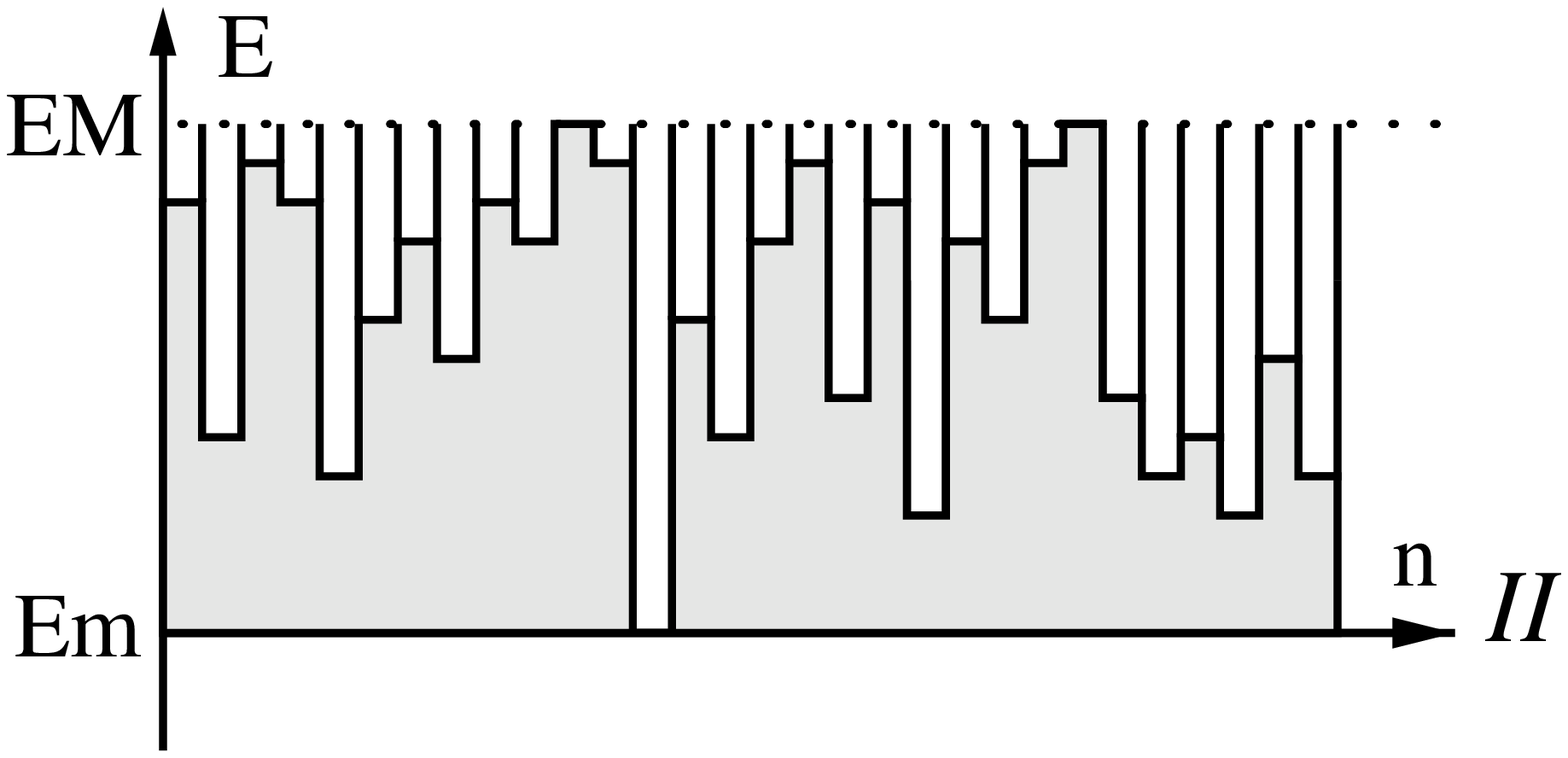}\\
      \includegraphics[width=0.24\textwidth]{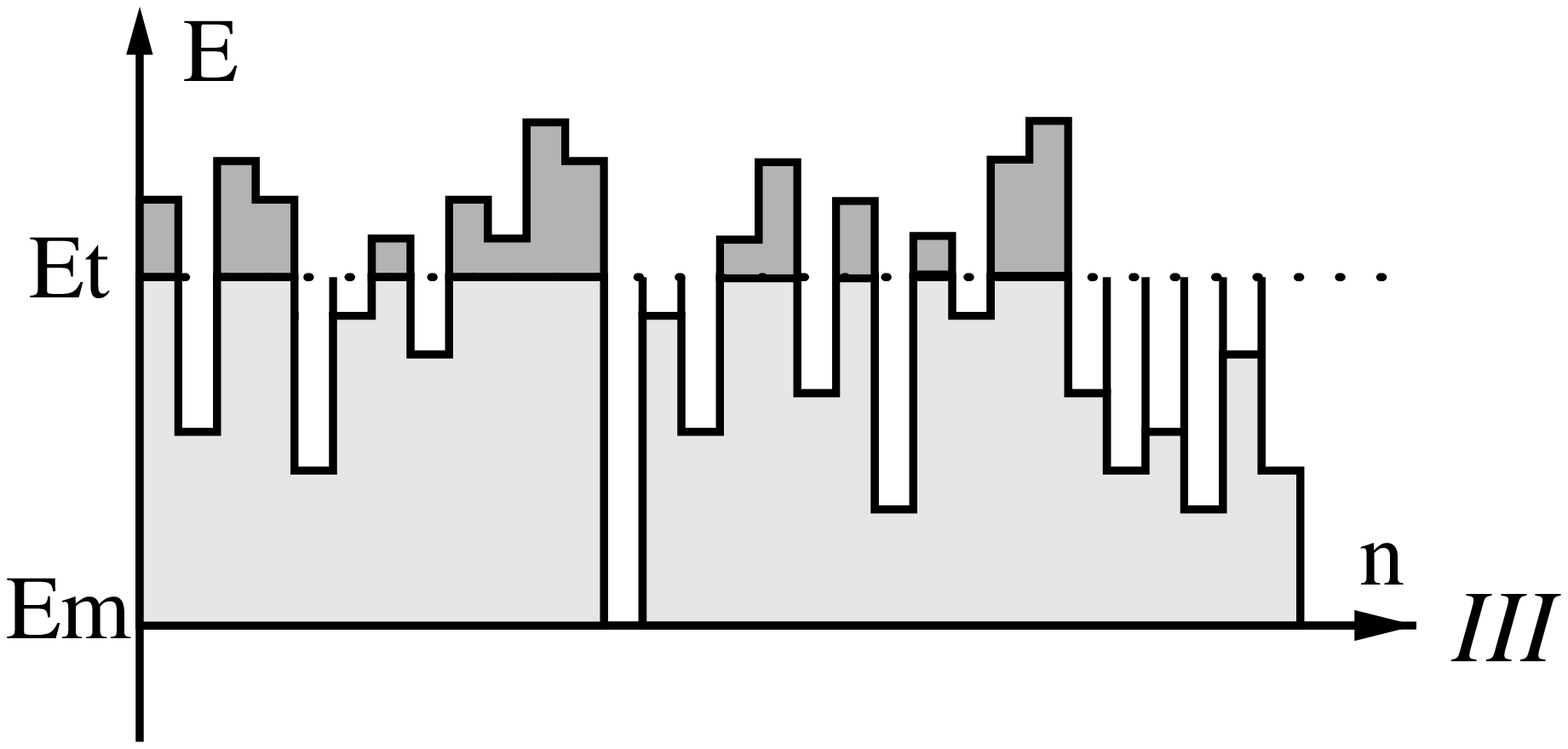} &
      \includegraphics[width=0.24\textwidth]{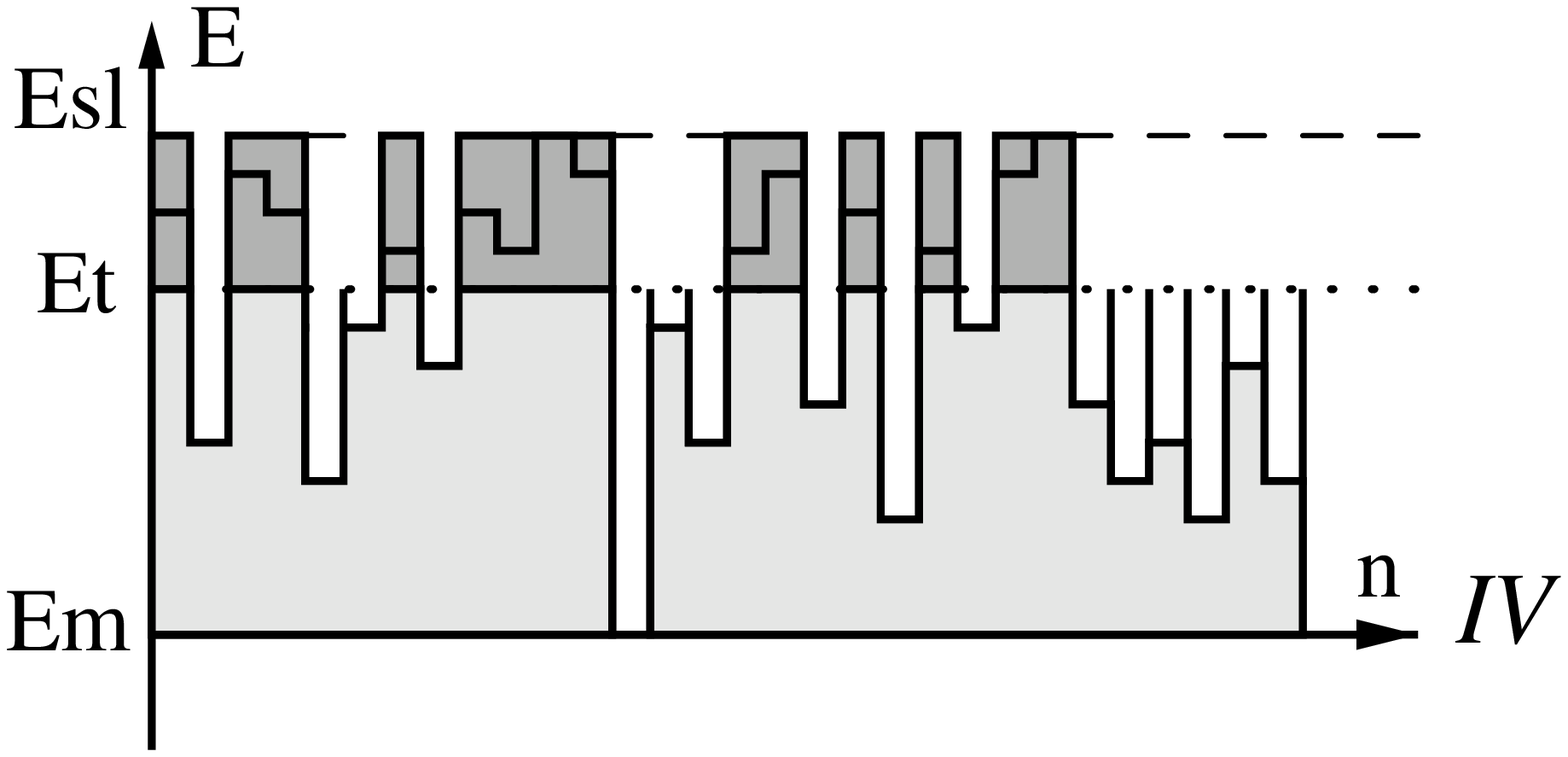}  \\
    \end{tabular}
    \caption{A schematic picture of the four considered variants
    of the model. On the horizontal axis, we represent a few (30)
    positions along DNA. Correspondingly we sketch the interaction
    energy $E$ varying between its minimum ($E_m$) and its maximum
    ($E_M$) values. The interaction energy evaluated on the T7 DNA
    present similar rapid oscillations between different levels. The
    dotted lines indicate the threshold level $E_t$, set to $E_M$ for model
    {\sl II}, to an intermediate value for model {\sl III} and {\sl
    IV}. In the case of model {\sl IV}, all energy levels above the
    threshold are redefined to a common value $E_{sl}$ (dashed line).}
    \label{fig:four}
  \end{center}
\end{figure}

To account for all these possibilities, we define and analyze some
different models, sketched in Fig.~\ref{fig:four} and listed
hereafter:
\begin{enumerate}

   \item[{{\em I)}}] {\em no-threshold model} (Fig.~\ref{fig:four},
   {\em I}): hydrogen bonds can directly translate from one position
   to another without being destroyed.  In this case the energy
   difference $\Delta E_{n \to n'}$  from $n$ to $n' = n \pm 1$ is
   simply
\begin{equation} \Delta E_{n \to n'} = \max [E(n')-E(n),0]\,; \end{equation} here $\Delta E_{n \to n'}$ is set to zero if $E(n')-E(n)$ is negative, as usual.

   \item[{{\em II)}}] {\em maximal-threshold
   model} (Fig.~\ref{fig:four}, {\em II}): in order to reach a next
   site, RNAP must destroy all bonds and pass through a state of
   ``total mismatch''. In this case $\Delta E_{n \to n'} = E_M-E(n)$,
   where $E_M=\max[E(n)]$.

   \item[{{\em III)}}] {\em intermediate-threshold
   model} (Fig.~\ref{fig:four}, {\em III}): in order to reach a next
   site, RNAP must destroy all bonds and pass through an intermediate
   ``zero'' state defined by a threshold energy $E_t$. One has
   therefore
\begin{eqnarray} \label{eq:threshold}
\Delta E_{n \to n'} =& 
 \max[E_t-& \hspace{-2mm} E(n), E(n')-E(n),0]\,.\nonumber
\end{eqnarray}
\end{enumerate}
Models {\em I}\, and {\em II}\, are actually the two limiting cases of
model {\em III}\, when the threshold is set to the minimum and maximum
values of the potential energy, respectively. These three models could
therefore be considered as three cases of a {\em unique model}, just
dependent on the choice of the energy threshold.  We will anyway refer
to these three cases as to models {\em I}\,, {\em II}\, and {\em
III}\, in the following, for convenience.  Note that in the general
case of an intermediate threshold, the previous model gives two
different possible regimes for the polymerase, because the energy
profile is qualitatively different in regions where $E(n)$ is greater
or lower than $E_t$.  

Finally, to account simultaneously for two possible regimes of the
RNAP-DNA interaction mentioned above, we propose a fourth model as
follows: \begin{enumerate}

   \item[{{\em IV)}}] {\em two-regimes model} (Fig.~\ref{fig:four}, {\em IV}): 
   a threshold energy  $E_t$ separates {\em``reading'' regions}, where the 
   energy is $E(n)<E_t$, from {\em ``sliding'' regions}, where no hydrogen bonds
   are made and the RNA polymerase can freely diffuse on a flat energy
   landscape, $E(n)=E_{sl}$. Below the threshold, the barrier $E_t$
   still affect the translocation as in case of model {\em III}.  For
   simplicity, we will fix the value of $E_{sl}$ to $E_M=\max[E(n)]$.
   In this case, one can redefine the energy as
   \begin{equation}
   \label{eq:4}
   E(n)=
   \left\{
   \begin{array}{ll}
   E(n) & \mbox{if } E(n) <   E_t\\
   E_{sl}  & \mbox{if } E(n) \geq E_t
   \end{array}
   \right.
   \end{equation}
   and  $\Delta E_{n \to n'}$  results to be defined as in case {\em
   III}\,.
  \end{enumerate}

Note that our model~{\em IV}\, interpolates between straight
sequence-dependent walk (model~{\em I}\,) and the biological model of
the promoter search proposed by von Hippel in
Ref.~\citep{von89,von96}.  The scenario suggested by von Hippel relies
indeed on the idea that the specific interaction is ``switched off''
by a conformational change if too many mismatches are present. In that
picture, RNAP is more often in a ``sliding'' mode, where the specific
hydrogen bond interaction is inactive. A quantitative description of
this mechanism can be obtained by the introduction of our model~{\em
IV}, where the varying threshold level $E_{t}$ accounts for the degree
of homology which leads to the supposed RNAP conformational change.

The rates $r_{n\to n'}$ of translocation between neighboring sites $n$
and $n'$ are, according to the Arrhenius law, proportional to $\exp{(-
\Delta E_{n \to n'} / k_B T)}$, where $n' = n \pm 1$.  The model
includes a nonzero probability for the polymerase to stop at one
position; the complete set of translocation rates reads therefore:
\begin{equation}
  \label{eq:rate}
\left\{
\begin{array}{lll}
r_{n\to n'}
&=& 1/2\, \exp{(- \Delta E_{n \to n'}/ k_B T)},\\
&&\hspace{2.5cm} n' = n \pm 1  \\
r_{n\to n}
&=& 1 -  r_{n\to n+1} - r_{n\to n-1}  \,.
\end{array} \right.
\end{equation}
In the case of flat energy landscape ($\Delta E_{n \to n'}=0$) all the
rates $r_{n\to n'}$ are equal to $1/2$, which defines a simple
one-dimensional diffusion process with diffusion constant $D = 1$.

If the discretization of length, $x=\ell n$ ($\ell=3.4 \; {\AA} $ is
the base pair step), and time, $t=\tau m$, is explicitly taken into
account, then the dimensionless $D$ given by the relation $\langle n^2
\rangle = 2D\; m$ corresponds to a physical value of
$D\ell^2/\tau$.  In order to give a quantitative meaning to our
results we need an estimate for the (mean) time $\tau$ required for
each translocation step.  The upper diffusion limit, $D = 1$,
associated to a physical diffusion constant $\ell^2/2\tau$, would
correspond to a free diffusion without any local trapping
effect. Schurr~\citep{Sch79} has estimated this upper limit of the
one-dimensional diffusion constant of lac repressor sliding and
rotating along DNA helix track to be $D_{lac} = 4.5 \; 10^{-9}
\; cm^2/s$. The lac repressor was approximated by a hard ball of radius
$a$ moving in a viscous medium.  Using the Schurr's approach, and
accounting for the difference in sizes between the lac repressor
$a_{lac}=4.9 \; 10^{-7} \;cm$ and the T7 RNA polymerase $a_{RNAP}
\approx 7 \; 10^{-7}\; cm$, the upper limit of the polymerase
diffusion constant would rescale as (see Ref.~\citep{Sch79} for
details): $D_{RNAP}=D_{lac}\,(a_{lac}/a_{RNAP})^3 \approx 1.54
\;10^{-9}\,cm^2/s$. The latter being compared to a ``free diffusion''
limit $l^2/(2\tau)$, $l=0.34 \;nm$, sets the elementary time interval
$\tau=l^2/(2D)\approx 3.8 \; 10^{-7}\;\mbox{ s}$, during which a
translocation to the nearest base pair may happen.

\begin{figure}[htbp]
  \begin{center}
   \includegraphics[height=0.48\textwidth,angle=-90]{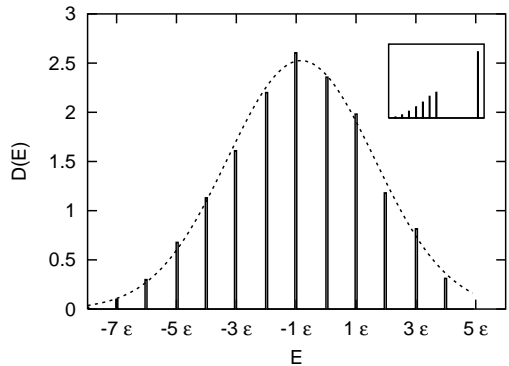}
    \caption{Energy level distribution (models I to III),
    obtained by averaging on the whole T7 DNA. A Gaussian fit of the
    resulting histogram ({\em dashed line}) is superimposed for
    comparison. {\em Inset:} The corresponding distribution for model
    IV ($E_t=0$).}
    \label{fig:levels}
  \end{center}
\end{figure}

Let us finally consider the distribution of energy levels that is
obtained when the real T7 DNA sequence is considered and the energy
landscape is evaluated through the local degree of homology by
Equation~(\ref{eq:energy}). In Fig.~\ref{fig:levels} the energy
distribution evaluated on the whole T7 sequence is represented. As can
be seen by comparing with the superimposed fit, the resulting
distribution  for models {\em I}\, to {\em III}\, is almost
Gaussian. Note that in the case of model {\em IV}\, all contributions
to levels above the threshold $E_t$ are obviously condensed in a
unique level $E_{sl}$ (see {\em inset} of Fig.~\ref{fig:levels}).

%%%%%%%%%%%%%%%%%%%%%%%%%%%%%%%%%%%%%%%%%%%%%%%%%%%%%%%%%%%%%%%%%%%%
\section{Results: recognition efficiency and anomalous diffusion}
\label{results}

The first important check of the four RNAP models is related to their
affinity to the promoter region.  Theoretically, one can easily
estimate the stationary distribution of a population of polymerase on
the four different model landscapes as
\begin{equation}
  \label{eq:rho}
  \rho_{_{\infty}}(n) \propto e^{-E(n)/ k_B T}
\end{equation}

As usual, the stationary distribution only depends on the site energy,
and not on differences and thresholds. Consequently, models {\em I}\,
to {\em III}\, have the same distribution, whereas the redefinition of
energy in model {\em IV}\, leads to a substantially different result.
Equation~(\ref{eq:rho}) straightforwardly implies that the recognition
sites, which have the lower energy, will be in average the most
populated.

In order to verify that this is indeed obtained in a dynamical
context, we simulated numerically the time evolution of models {\em
I}\, to {\em IV}\, taking a uniform distribution of independent RNAPs
on a DNA region of $1000$ bps as initial condition.  Note that the
assumption of an uniform initial distribution is statistically
equivalent to considering the probability evolution of a single
polymerase binding to DNA at random site.  The simulation is performed
on the first $3000$ base-pairs of the T7 sequence, which contains two
recognition sequences GAGTC, at positions $1126$ and $1435$.

\begin{figure}[htbp]
  \begin{center}
    \includegraphics[height=0.48\textwidth, angle=-90]{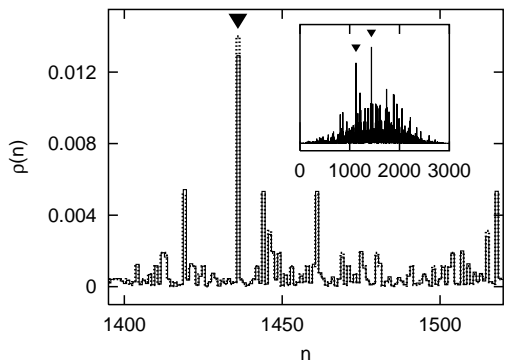}
    \caption{A central portion of the polymerase distribution
     $\rho(n)$ for model {\em I}\, after an integration time of $10^6$
     integration steps, obtained by averaging over $3 \, 10^4$
     particles initially uniformly distributed in the interval
     $[1000,2000]$ ({\em solid line}).  The analytical equilibrium
     distribution $\rho_{_\infty}(n)$, ({\em dotted line}) is shown
     for comparison. Here ${\mathcal \epsilon}/ k_B T = 0.5$.  {\em
     Inset:} the whole distribution at the same time.  In both plots,
     the arrows indicate the location of the recognition sequences
     GAGTC (sites 1126 and 1435). }
     \label{fig:distr}
  \end{center}
\end{figure}

After a sufficiently long time, the polymerase distribution $\rho(n)$
spreads out, as shown in the inset of Fig.~\ref{fig:distr}, and shows
a series of peaks corresponding to the sites with larger occupancy.
Where the border effects can be neglected, this distribution tends to
its equilibrium limit; this is shown in Fig.~\ref{fig:distr}, where we
plot a portion of the distribution obtained after $10^6$ time steps
for model {\em I}\,, together with $\rho_{_\infty}$. As expected, the
larger peaks correspond to energy minima, i.e., to the location of the
two recognition sequences GAGTC present in this DNA region.  For all
the models {\em I}\, to {\em III}\, the final distribution is similar,
with the two highest peaks exactly in correspondence to the two
recognition sequences, this confirming that the energy landscape
defined on the basis of the pattern matching actually guides the
polymerase to the promoter recognition sequences.

Note that, in case of model {\em IV}\,, the distribution of levels
is different, this obviously implying a different shape for
$\rho_{_\infty}(n)$. The case of a sufficiently low threshold energy is
reflected on an asymptotic distribution with rarer, larger peaks
on a very low constant background (data not shown).

We now investigate the dynamical behavior of the four models, and
check if there are some relevant deviations from random walk, induced
by the sequence sensitivity. For large enough values of ${\mathcal
\epsilon}/ k_B T$, some positions along DNA could trap polymerase for
long time, this implying that, at small and intermediate time,
diffusion could be substantially different than for a pure random
walk. In order to estimate this effect, we calculate the mean square
displacement for the polymerase: 
\begin{equation}
  \label{eq:deltan2}
\langle \Delta n ^2\rangle =\langle \Delta n ^2 (t)\rangle
= \sum_{i=1}^N (n_i(t)-n_i(0))^2 \,.
\end{equation}
We average over $N=9\, 10^3$ independent particles, initially
distributed uniformly in the DNA region $[1000,2000]$.  This procedure
therefore includes both average on a large number of particles and on
a large set of initial conditions.

\begin{figure}[htbp]
  \begin{center}
    \includegraphics[height=0.48\textwidth, angle=-90]{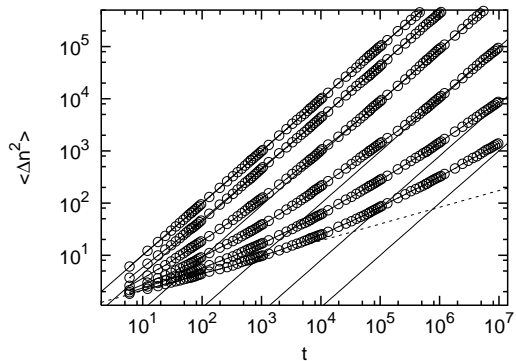}
    \caption{Diffusion behavior of model {\em I}\, for different
    values of ${\mathcal \epsilon} / k_B T$. From the upper curve to the
    bottom: $ {\mathcal \epsilon}/ k_B T =
    0,\,0.3,\,0.6,\,0.9,\,1.2,\,1.5$. Note the log-log scale: a linear
    diffusion $ \langle \Delta n ^2\rangle \propto t$ corresponds in
    this graph to the straight lines of unit slope ({\em solid
    lines}), while slopes lower than $1$ correspond to $ \langle
    \Delta n ^2\rangle = A\, t^b$, with $b<1$. A ({\em dashed}) line
    of slope $0.3$ is reported for comparison.}
    \label{fig:DbetaE}
  \end{center}
\end{figure}

Starting from model {\em I}\,, we investigate the dependence of the
diffusive behavior on ${\mathcal \epsilon}/ k_B T$. Results are shown in
Fig.~\ref{fig:DbetaE}. In the limit of ${\mathcal \epsilon}/ k_B T = 0$,
i.e., in the case of a flat potential (or $T=\infty$), the diffusion is
of course normal, with $D=1$ and $\langle \Delta n ^2 (t)\rangle =
2t$, so that the corresponding curve is a straight line of slope 1 in
the log-log plot (upper curve on Fig.~\ref{fig:DbetaE}).  For larger
values of ${\mathcal \epsilon} / k_B T$ (smaller temperatures compared
with the energy fluctuations), the dynamics of the model shows
initially large deviations from the normal diffusion: in these finite
temperature cases, the motion is initially subdiffusive, with
\begin{equation}
 \langle \Delta n ^2\rangle = A\, t^b\,, \,\,\,b<1 \,.
\end{equation}
The exponent $b$ increases monotonically with time towards its
asymptotic value $1$. The initial deviation $(1-b)$ and the crossover
to $b=1$ both increase with ${\mathcal \epsilon} / k_B T$.  This
behavior does not depend on the choice of the initial condition and
it is not a transient induced by some $t=0$ properties: we have
verified indeed that qualitatively the same time dependence is
reproduced after an initial transient time of $10^4$, $10^5$ or $10^6$
time steps.  As expected, once reached the normal diffusion regime,
different temperatures correspond to different diffusion constants $D$
(in the log-log representation, $2D$ corresponds to the vertical
offset of the lines of slope $1$, according to the relation $\log
\langle \Delta n ^2\rangle = \log 2D + \log t$).

Plots of Fig.~\ref{fig:DbetaE} also give a measure of the slowing down
in the promoter search induced by the sequence-dependent interaction.
Indeed, in the log-log plot the horizontal offset, at a given 
$\Delta n^2$, between different curves corresponds to the logarithm of the
ratio between the time needed to cross the corresponding displacement
$ {\Delta n}$ for different choices of ${\mathcal
\epsilon}/ k_B T$. Therefore, if $ {\Delta n}$ is a typical 
distance to promoter, the horizontal offset just gives the slowing
factor induced by subdiffusion with respect to normal
diffusion. Referring to Fig.~\ref{fig:DbetaE}, we can conclude that,
if the distance to promoter is larger than $100 \,bps$ (so that
${\Delta n}^2 = 10^4$), then the time to reach the promoter should be
reduced with respect to standard diffusion roughly of a factor $10$
for the case ${\mathcal \epsilon}/ k_B T = 0.6$, of a factor $100$ for
${\mathcal \epsilon}/ k_B T = 0.9$.  Furthermore, this slowing factor
does not depend on ${\Delta n}$, provided that it is large enough to
consider the asymptotic regime.  In this hypothesis, it is possible to
obtain an analytical estimation of the slowing factor \citep{forthcoming}.

\begin{figure}[htbp]
  \begin{center}
    \includegraphics[height=0.48\textwidth, angle=-90]{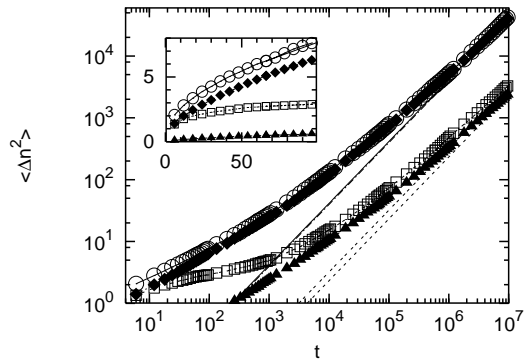}
    \caption{Mean square deviation $\langle \Delta n ^2 \rangle$
    for the four different models, with ${\mathcal \epsilon}/ k_B T =1$
    and $E_t=0$, in the log-log representation. Symbols refer
    respectively to: open circles, model {\em I}\,; triangles, model
    {\em II}\,; diamonds, model {\em III}\,; squares, model {\em IV}\,
    ($E_t=0$). The straight lines correspond to the fit in the last
    part of the graphs ($t \in [6\,10^6, 10^7]$). {\em Inset:} the
    same curves in a linear representation in the short time regime
    (symbols have the same meaning).}
    \label{fig:diff}
  \end{center}
\end{figure}

We will now extend the diffusion analysis to the other versions of the
model, introduced in Section~\ref{methods}.  Resulting curves for
models {\em I}\, to {\em IV}\, and for ${\mathcal \epsilon}/k_B T=1$ are
presented in Fig.~\ref{fig:diff}.  As for model {\em I}\,, in all
cases we observe at short time a subdiffusive regime due to the
trapping effect of the rough energy landscape.

The initial values of $b$, fitted in the time range $(0,100)$ through
the function $A \,t^b$, are the following for the first three
models:
\begin{eqnarray}
I:\, b= 0.49 \pm 1 \% \nonumber \\
II:\, b= 0.61 \pm 1 \%\nonumber \\
III:\, b= 0.56 \pm 1 \%\nonumber \,.
%\\IV:\, b= 0.25 \pm 4 \%\,.
\end{eqnarray}
Note that model {\em IV} displays in this short time regime a
particular behavior, that will be discussed in the following.

Let us remark that, in principle, the obtained anomalous diffusion
could be due to some particular spatial correlation properties of the
underlying potential.  Nevertheless, we have checked that it is only
due to the roughness of the landscape, doing the same experiment on an
artificial base sequence, completely random. In the conditions
described by model {\em I}\,, for instance, and in the same fit range,
we obtained $b=0.52\, \pm \, 1\%$, and a curve similar to T7 DNA case
(data not shown).

\begin{figure}[htbp]
  \begin{center}
    \includegraphics[height=0.48\textwidth, angle=-90]{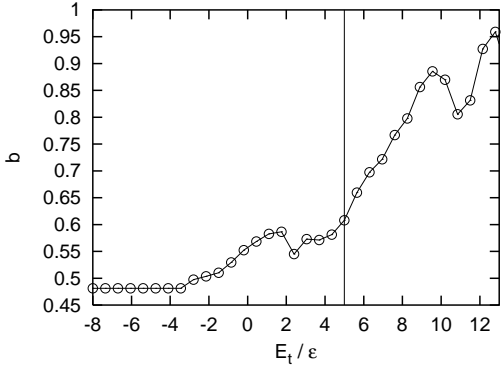}
    \caption{Behavior of the exponent $b$ as fitted in the short
    time regime $t \in (0,100)$ as a function of the threshold energy
    $E_t$ for model {\em III}. The vertical line corresponds to
    $\max[E(n)] = 5 \mathcal \epsilon$.}
    \label{fig:bshort}
  \end{center}
\end{figure}

We then studied the behavior of the short time subdiffusive exponent
$b$ as a function of $E_t$ for model {\em III} with varying threshold
(i.e., including model {\em I} and {\em II}\,). The results are shown
on Fig.~\ref{fig:bshort}. For threshold lower that a critical value of
about $-3 \mathcal \epsilon$ the system displays almost no sensitivity
to the threshold level. Indeed, This is due to the fact that, some
relevant effect, it is necessary to have not only a site $n$ with
$E(n)<E_t$, but also at least two neighboring sites should be below
the threshold in order to feel its effect (see
Eq.~\ref{eq:threshold}). The probability of finding two adjacent sites
below the threshold is too low below $E_t\leq -3 \mathcal \epsilon$,
thus explaining the observed insensitivity.  Interestingly, the
exponent $b$ becomes a nonmonotonic and very sensitive function of
$E_t$ for larger values of $E_t$. The effect of the threshold in this
intermediate regime is in fact twofold: from one side, it induces an
additional damping on many low energy sites; from the other, it makes
(a fraction of) these same sites ``blind'' to the energies of their
neighborings (the translocation barriers only will depend on $E(n)$
and $E_t$). The complex balance between the two contributions induces
the high instability of the fit results displayed in
Fig.~\ref{fig:bshort}.  As the threshold increases above the maximum
level ($E_t=5 \mathcal
\epsilon$), the disorder of the underlying energy landscape becomes
less and less important, and the system tends to recover a standard
diffusive behavior strongly damped, i.e., with $b \to 1$ and $A \to 0$.

Now let us consider the large time limit.  The asymptotic diffusion
constant depends on the model choice.  A linear fit of the large time
regime of $\langle \Delta n ^2 \rangle$ of Fig.~\ref{fig:diff} has
been done in order to estimate the average diffusion constant $D$, in
the random walk approximation where $\langle \Delta n ^2\rangle =
2Dt$. Besides, we checked that an effective linear behavior is reached
in the corresponding time range by fitting again with a function
$\langle \Delta n ^2\rangle = A\,t^b$ and verifying that $b$ is close
to unity.  The resulting diffusion constants $D$ and the exponents $b$
for the four models at large time ($t\in [6\,10^6, 10^7]$)
are given, for $E_t=0$, respectively by:
\begin{eqnarray}
\label{long}
I:\,&2D=\, 4.1\; \,10^{-3} \pm 1 \% &\;\;\;\;\; b=0.93 \pm 1 \% \nonumber \\
II:\,&2D=\,0.23\,10^{-3} \pm 2 \% &\;\;\;\;\; b= 0.86\pm 1 \%\nonumber \\
III:\,&2D=\, 4.0\; \,10^{-3} \pm 1 \% &\;\;\;\;\; b= 0.91\pm 1 \%\nonumber \\
IV:\,&2D=\, 0.32\,10^{-3} \pm 1 \% &\;\;\;\;\; b= 0.85 \pm 1 \%\,.
\end{eqnarray}
The corresponding fits are the straight lines in Fig.~\ref{fig:diff}.

\begin{figure}[htbp]
  \begin{center}
    \includegraphics[height=0.48\textwidth, angle=-90]{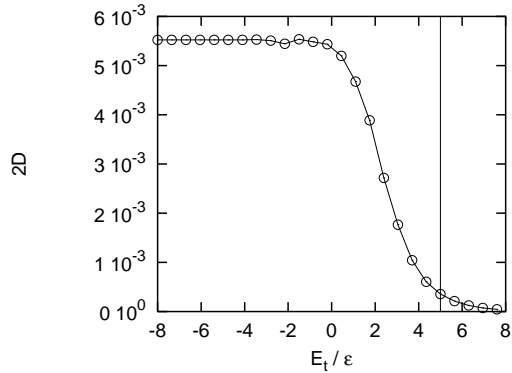}
    \caption{Behavior of the coefficient $2D$ as fitted in the
    large time regime $t \in (8\, 10^5, 10^6)$ as a function of the
    threshold energy $E_t$ for model {\em III}. The level $\max[E(n)]
    = 5 \mathcal \epsilon$ is represented by a vertical line.}
    \label{fig:2DEt}
  \end{center}
\end{figure}

The differences in the equilibrium diffusion constant between
different models are explicitly related to the activation barrier in
the four cases: the higher is the threshold to overcome in order to
move one step, the lower is the diffusion constant.  Note that in the
case of model {\em IV}\, the boundaries between flat and rough regions
act as energy barriers of amplitude $\approx E_{sl}$: these barriers
appear to affect the motion more strongly than the threshold $E_t$,
this resulting in a diffusion constant closer to that of model {\em
II}\, than to that of model {\em III}\,.

We have analyzed the dependence on $E_t$ also for the asymptotic
diffusion constant $D$. Fig.~\ref{fig:2DEt} shows the dependence of
$D$ on $E_t$ in model {\em III}\,\footnote{For technical reasons, we
display data resulting from the fit in the range $(8\, 10^5, 10^6)$,
i.e., in a region where the parameter $b$ has not yet reached
unity. The curve of Fig.~\ref{fig:2DEt} represents therefore only a
qualitative analysis and shows some small discrepancy with data given
in Eq.~\ref{long}.}.  Again, almost no sensitivity to the threshold
level is observed below a critical value, approximatively $E_t=-3
\mathcal \epsilon$. Roughly, between this value and $E_t=0$, we observe
a transition to a regime of strong sensitivity ($E_t>0$), where the
damping effect induced by the threshold is much more enhanced. The
diffusion constant decreases rapidly above the maximal energy ($E_M=5
\mathcal \epsilon$, vertical line), as intuitively expected.

\begin{figure}[htbp]
  \begin{center}
    \includegraphics[height=0.48\textwidth, angle=-90]{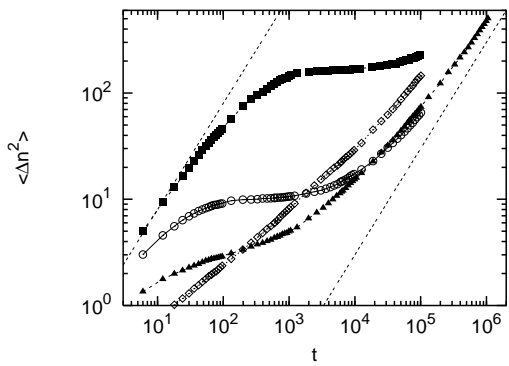}
    \caption{Time behavior of $\langle \Delta n ^2\rangle$ for
    model~{\em IV}\,, in the cases $E_t=-4$ ({\em full squares}),
    $E_t=-2$ ({\em circles}), $E_t=0$ ({\em full triangles}), $E_t=2$
    ({\em diamonds}), and with ${\mathcal \epsilon}/ k_B T = 1$. Two
    straight lines of slope $1$ are shown for comparison.}
    \label{fig:fourEt}
  \end{center}
\end{figure}

We shall now discuss in detail model~{\em IV}\,, since it displays,
with respect to the others, a more complicated behavior.  Note that,
in principle, model~{\em IV}\, can be put exactly in the same scheme
as the other models, once the underlined potential $E(n)$ is redefined
according to Equation~(\ref{eq:4}). Nevertheless, this redefinition of
the energy landscape leads to substantially different features. As can
be observed in Fig.~\ref{fig:fourEt}, during an initial time interval
the polymerase diffuses more rapidly, even if still subdiffusively,
with initially a larger effective diffusion constant.  The initial
speeding up of the dynamics becomes more pronounced as the value of
the threshold decreases, i.e., as the energy redefinition involves an
increasing number of sites.  This effect can be explained by
considering how the potential landscape is changed for model~{\em
IV}\,.  Among the particles, uniformly distributed at time zero over a
large region of the sequence, all those that are initially on flat
regions of energy $E_{sl}$ will start diffusing freely with diffusion
constant equal to $1$, until they fall down in one $E<E_t$
region. These particles contribute initially to the diffusion with a
large term, thus making it increase.  After an initial transient,
however, most of the particles will be almost trapped in the potential
wells, and the effective diffusion coefficient will decrease
accordingly.

More precisely, the trapping effect will depend on the value of
$E_{sl}$, set to $\max{[E(n)]}$ in our calculations. If $E_{sl}$ is
big enough, most of the particles will be trapped in $E(n)<E_t$
regions, with activation barriers and only a small probability to
escape again toward the flat plateaux.  Therefore, in the long time
regime, the system will be essentially in the same state model~{\em
III}\,, but mostly localized in some finite regions.  In other words,
the particular equilibrium conditions introduced in model {\em IV} are
indeed such that one particle needs to spend a large amount of energy
(and, therefore, of time) before reaching a high level plateau, but
once reached, it can move much faster to the next favorable site.  An
analytical derivation of the main dynamical quantities as functions of
the model parameters discussed in this section will be presented
elsewhere \citep{forthcoming}.

%%%%%%%%%%%%%%%%%%%%%%%%%%%%%%%%%%%%%%%%%%%%%%%%%%%%%%%%%%%%%%%%%%%%
\section{Discussion}
\label{comments}

All the results presented in this work can be checked by a comparison
with detailed experimental data. As mentioned in the introduction,
experiments leading to a rather precise determination of the RNAP
position along DNA at different times during the promoter search have
already appeared \citep{Kab93,Har99,Gut99}, and others are in progress
\citep{Place}. This will give for the first time the possibility to
estimate the detailed features of the T7 RNAP diffusive motion.  As we
have shown, a dynamical model which includes both the affinity for the
promoter together and the possibility of sliding, leads to a
nontrivial sequence dependent dynamics, at least in some range of the
parameters. It is thus important to verify if these effects can
actually be observed experimentally.  The sliding distance is
kinetically evaluated in different experiments around $350-1000$ bps
(\citep{Shi99} and references therein). This is probably not peculiar
to RNAP since other enzymes also seem to slide along the DNA covering
a short distance of about $300$ bps before being released in solution
\citep{Sta00}. In this space scale, the anomalous diffusion behavior
is predominant for our model.

In particular, the recent scanning force microscope (SFM) experiment,
performed by Gut\-hold {\it et~al.} \citep{Gut99}, allows for a direct
observation of one {\it E. coli} RNAP sliding back and forth on a single
DNA chain partially adsorbed on a mica surface, although with some
technical limitations (the average lifetime of the nonspecific complex
is more than hundred times larger than what measured in solution,
probably due to the two-dimensional constraints).  The statistical
properties of the observed diffusive motion have been fitted by the
law $\langle\Delta x^2 \rangle = 2D\,t$, in order to confirm the
general assumption that RNAP moves randomly along DNA (\citep{Gut99},
Fig.~2).  Quantitatively, however, in the observed displacement ranges
(less than two hundreds base-pairs), the corresponding data seem to
deviate from a pure diffusive motion. This may be due to the
experimental constraints and to the limited number of RNAP sliding
trajectories (about 30). On the other hand, the rough estimate of
numerical data from Fig.~2 of Ref.~\citep{Gut99}, fitted with a power
law of the type $A t^b$, gives $b \sim 0.5 \pm 15 \%$. It is very
interesting to note that these data seem much more compatible with a
subdiffusive behavior than with normal diffusion, as is usually
assumed.  This first experiment allowing for a direct visualization of
the RNAP sliding motion gives therefore, from our point of view,
intriguing and encouraging results.

We remark that the dynamical features described here depend crucially
on the choice of the model parameters: the ratio ${\mathcal \epsilon}/
k_B T$, the value of the energy threshold $E_t$, and, in the case of
model {\em IV}\,, the energy of the plateaux $E_{sl}$.  As a first
check, we can try to compare our rough estimation of the power
exponent we estrapolate from the results in Ref.~\citep{Gut99} with
the behavior of the model as a function of ${\mathcal \epsilon} / k_B
T$.  The value of about $0.5$ very roughly corresponds to ${\mathcal
\epsilon} / k_B T \approx 1$ for all values of $E_t$, this confirming
that the parameter choice made in the most part of our simulations
could be indeed of the right order of magnitude.

Further experimental investigations, devoted to the detailed
determination of the nonspecific interaction, are necessary to improve
the model.  The version of the model which is compatible with the
sliding RNAP dynamics of single molecule experiments should emerge
from comparison with the experimental data, using the model parameters
as fitting parameters. In practice, the complicated diffusive behavior
of the model will allow us to compare theory and experiments by means
of more than one dynamical observable. For the case of model {\em
IV}\,, where the additional model parameter $E_{sl}$ is needed, the
presence of a new short-time specific feature could be used in the fit
of the experimental results.

From a biological point of view, the four models offer a framework
for defining the pertinent parameters to optimize the promoter
search. For all models, the specific interaction energy $\mathcal
\epsilon$ between RNAP and DNA is crucial and should be close to
$k_BT$ in order to allow the polymerase to move. This adjustment of
the interaction energy can be achieved by varying the distance and
angle of the H-bonds during sliding. Perhaps, the more interesting
model from a biological point of view is model {\em IV\,}, since it
allows for a better control of the diffusion pattern, and consequently
for the corresponding biological function. An exact balance has to be
found in biological system between the reading and sliding
mode. $E_t$, $E_{sl}$, and ${\mathcal \epsilon} / k_B T$ have to be
optimized for the biological purpose which will be physically
reflected by the protein-DNA interaction and by the DNA sequence.

Finally, it is important to keep in mind that the recognition
mechanism through hydrogen bonds considered here does not allow for a
complete identification of the promoters. The recognition sequence
GACTC (or of the complementary sequence GAGTC) appears in the complete
T7 genome more than 90 times; however, only 10 of them actually
belongs to 17 bps long promoters.  Evidently, other ``signals''
cooperate with the direct pattern recognition mechanism in order to
allow the polymerase to find its target.  The weak sequence TAATA
(positions -13 to -17), for instance, also interacts with RNAP through
the minor groove \citep{Che99}.  A sensitivity to this minor groove
region should probably be included. In this sense, our model
represents a first attempt towards a detailed description of the RNAP
dynamics during the promoter search. The model can also be extended to
the case of other enzymes by a detailed introduction of their
sequence-dependent interaction with nonspecific DNA. We believe that
the main idea of the model, which is the link between base sequence
and enzyme dynamics, will be valid in general.  Indeed, as far as a
sequence dependence is considered, the enzyme will always interact
with DNA through an effective potential with a fluctuating
profile. This potential should be induced for different enzymes by
different kinds of interaction. Its roughness by itself, however, will
always generate anomalous diffusion features as those described in
this paper.

\section*{Conclusions}
\label{conclusion}

In this paper we have proposed a simple model for the RNAP sliding
motion along DNA, which includes a sequence dependent interaction.  We
deduced an hypothetical polymerase-DNA interaction from the
crystallographic structure of the T7 polymerase-promoter complex
\citep{Che99}.  We have included four possible variations by
considering slightly different translocation probabilities, i.e.,  by
the presence of a variable activation barrier $E_t$ (leading to models
{\em I}\, to {\em III}\,), and eventually by distinguishing
{\em``reading''} regions from {\em ``sliding''} regions, where no
hydrogen bonds are made so that the RNAP can freely diffuse on an
effective constant potential (model~{\em IV}\,).

A numerical study of the diffusion properties of the four versions of
the model shows that a normal diffusion regime is only achieved after
some time.  We have shown as all the four models are characterized at
shorter times by a subdiffusive behavior. A rough estimation of the
slowing factor induced by the sequence dependence for different values
of the energy parameter can be easily obtained. This result is of
particular interest because, as we have discussed, the anomalous
diffusion is observed in a range that corresponds approximatively to
the experimentally observed characteristic distance covered by the
RNAP during sliding \citep{Shi99}.  The physical reasons underlying
the different diffusion behaviors have been discussed.

Nowadays the existing nano-technologies and single molecule techniques
allow for constraining and manipulating single biological objects.
The present paper represents a first step towards theoretical picture
where some of the resulting experimental results could be analyzed and
connected with the known functional properties of the corresponding
biological systems. It is important to keep in mind, anyway, that the
in vivo dynamics of the corresponding biological processes
occurs in a high density environment, in presence of very complex
spatial structures and of water molecules mainly bound and structured
\citep{Goo92}. What we usually call the diffusive motion of proteins
inside the cell is likely to be instead a motion strongly dependent on
a complex set of environmental trapping sites, as in the case
considered here. Also in this respect, the approach proposed in this
paper may have a larger range of application.

\nocite{Gue02}

\vskip 1cm  %\centerline{ {\bf{ACKNOWLEDGMENTS}}} \vskip .5cm
We are grateful to  A. Lesne, M. Peyrard  and S. Ruffo for helpful
discussions.  M.B. wishes to thanks the EU and the Physics Department of
the University of Salerno, Italy, for a two years post-doctoral research
grant during which this work was done. V.P. acknowledges the Physics
Department of the University of Salerno for financial support of two
short term visits during which part of this work was done. M.S.
acknowledges partial support from MURST through a PRIN-2000 Initiative
and from the European grant LOCNET n.o HPRN-CT-1999-00163.

\bibliographystyle{apsrmp2}
\bibliography{bibrnap}

%%%%%%%%%%%%%%%%%%%%%%%%%%%%%%%%%%%%%%%%%%%%%%%%%%%%%%%%%%%%%%%%%%%%%%
\end{document}